\documentclass[prl,reprint,longbibliography,superscriptaddress]{revtex4-2}
\usepackage{graphicx}
\usepackage{amssymb,amsmath,amsthm,bm}
\usepackage{multirow} 
\usepackage{physics}
\usepackage{braket}
\usepackage{times}
\usepackage{lipsum}
\usepackage{gensymb}
\usepackage{color}
\usepackage{pifont}
\usepackage{float}
\usepackage{xcolor}
\usepackage{booktabs}

\usepackage[unicode=true, bookmarks=true,bookmarksnumbered=false,bookmarksopen=false, breaklinks=false,pdfborder={0 0 1},backref=false,colorlinks=true]{hyperref}

\hypersetup{linkcolor=magenta,urlcolor=blue,citecolor=blue,pdfstartview={FitH},hyperfootnotes=false}

\usepackage[export]{adjustbox}

\begin{document}
\title{Beating the break-even point with autonomous quantum error correction}

\author{Yi Li$^{1,2,5,6}$}
\thanks{These authors contribute equally to this work.}
\author{Qingyuan Mei$^{1,2}$}
\thanks{These authors contribute equally to this work.}
\author{Qing-Xuan Jie$^{3,4}$}
\thanks{These authors contribute equally to this work.}
\author{Weizhou Cai$^{3,4}$}
\author{Yue Li$^{1,2}$}
\author{Zhiyuan Liu$^{1,2}$}
\author{Zi-Jie Chen$^{3,4}$}
\author{Zihan Xie$^{1,2,6}$}
\author{Xu Cheng$^{1,2,6}$}
\author{Xingyu Zhao$^{1,2,6}$}
\author{Zhenghao Luo$^{1,2}$}
\author{Mengxiang Zhang$^{7}$}
\author{Xu-Bo Zou$^{3,4}$}
\author{Chang-Ling Zou$^{3,4,6}$} \email{clzou321@ustc.edu.cn}
\author{Yiheng Lin$^{1,2,6}$} \email{yiheng@ustc.edu.cn}
\author{Jiangfeng Du$^{1,2,6,8}$} 
\email{djf@ustc.edu.cn}

\maketitle

\onecolumngrid
\begin{center}
\small
\begin{minipage}{0.75\textwidth}  
\centering
\vspace{-3.5em}
\noindent
\textit{
$^1$Laboratory of Spin Magnetic Resonance, School of Physical Sciences, University of Science and Technology of China, Hefei 230026, China\\
$^2$Anhui Province Key Laboratory of Scientific Instrument Development and Application, University of Science and Technology of China, Hefei 230026, China\\
$^3$CAS Key Laboratory of Quantum Information, University of Science and Technology of China, Hefei 230026, China.\\
$^4$Anhui Province Key Laboratory of Quantum Network, University of Science and Technology of China, Hefei 230026, China\\
$^5$National Advanced Talent Cultivation Center for Physics, University of Science and Technology of China, Hefei 230026, China\\
$^6$Hefei National Laboratory, University of Science and Technology of China, Hefei 230088, China\\
$^7$Anhui Province Engineering Research Center for Quantum Precision Measurement, Hefei 230088, China\\
$^8$Institute of Quantum Sensing and School of Physics, Zhejiang University, Hangzhou 310027, China}
\end{minipage}
\end{center}

\begin{center}
\begin{minipage}{0.8\textwidth}
Quantum error correction (QEC)~\cite{Nielsen2023, RevModPhys.87.307} is essential for practical quantum computing, as it protects fragile quantum information from errors by encoding it in high-dimensional Hilbert spaces. Conventional QEC protocols~\cite{KITAEV2003, Dennis2002} typically require repeated syndrome measurements, real-time feedback, and the use of multiple physical qubits for encoding. Such implementations pose significant technical complexities, particularly for trapped-ion systems, with high demands on precision and scalability. Here, we realize autonomous QEC~\cite{Reiter2017, may2020, Gertler2021, LiZQ2024a, Lachance-Quirion2024} with a logical qubit encoded in multiple internal spin states of a single trapped ion~\cite{Zhang2022, Ringbauer2022, Meth2025}, surpassing the break-even point for qubit lifetime.  Our approach leverages engineered spin-motion couplings to transfer error-induced entropy into motional modes, which are subsequently dissipated through sympathetic cooling with an ancilla ion, fully eliminating the need for measurement and feedback. By repetitively applying this autonomous QEC protocol under injected low-frequency noise, we extend the logical qubit lifetime to approximately 11.6\,ms, substantially outperforming lifetime for both the physical qubit  ($\simeq$0.9\,ms) and the uncorrected logical qubit ($\simeq$0.8\,ms), thereby  beating the break-even point with autonomous protection of quantum information without measurement~\cite{Ofek2016, Ni2023, Sivak2023, Acharya2025} or post-selection~\cite{Hong2024,Egan2021}. This work presents an efficient approach to fault-tolerant quantum computing that harnesses the intrinsic multi-level structure of trapped ions, providing a distinctive path toward scalable architectures and robust quantum memories with reduced overhead.

\end{minipage}
\end{center}

\twocolumngrid
\maketitle
\noindent \textbf{\large{}Introduction}{\large\par} 
\noindent Quantum computers promise substantial computational power for quantum chemistry, quantum many-body physics, and cryptography~\cite{alanaspuru-guzik2005, shor1997, google2020, bernstein2017}. However, the inherent coupling between quantum systems and environmental noise can lead to errors that corrupt quantum information, presenting a significant obstacle to practical quantum computing~\cite{Morvan2024, Sun2024}. Quantum error correction (QEC) techniques have been developed to suppress the effects of noise by encoding information into higher-dimensional Hilbert spaces, thereby identifying and correcting errors, reducing the impact of noise to higher orders~\cite{Nielsen2023, RevModPhys.87.307}. Experimental implementations of QEC have advanced significantly in recent years across multiple platforms, such as neutral atoms~\cite{Bluvstein2024, Scholl2023}, superconducting systems~\cite{Devoret2013}, nuclear spin systems~\cite{Lim2025}, and trapped ions~\cite{Egan2021, Erhard2021, Self2024, Pogorelov2025, Hong2024, Ryan-Anderson2024}. Notably, superconducting systems have achieved a crucial milestone by demonstrating logical qubit coherence times that surpass those of the best physical qubits, thereby reaching the break-even point of QEC~\cite{Ofek2016, Sivak2023, Ni2023,Acharya2025}.

Trapped ions are among the most advanced and versatile platforms for quantum computing, offering long coherence times~\cite{wang_single_2021}, high-fidelity quantum gates~\cite{gaebler_high-fidelity_2016, Ballance2016}, and precise control over both internal spin states and motional degrees of freedom~\cite{monroe_programmable_2021, chen_quantum_2021}. Despite these advantages, the technical demands of syndrome measurement and real-time feedback pose substantial challenges for implementing repetitive QEC protocols with trapped ions. Fortunately, trapped ions naturally possess rich multi-level internal electronic states (qudits) and external motional modes, enabling alternative hardware-efficient encoding schemes that can significantly reduce the physical resources required for QEC. Recently advances about the logical qubit encoding within single qudit~\cite{Kubischta2023, Kubischta2024, Gross2024, Lim2023, gross2021, Omanakuttan2024,yu_schrodinger_2025} or bosonic mode~\cite{Fluhmann2019,Cai2021}  have demonstrated these hardware-efficient advantages. Moreover, these rich internal and external degrees of freedom offer an elegant solution to the measurement and feedback challenge: employing engineered dissipative processes in trapped ions to autonomously correct errors without measurement~\cite{Reiter2017, may2020, Gertler2021, LiZQ2024a, Lachance-Quirion2024}. 

Here, we combine the advantages of multi-level encoding and autonomous error correction in a trapped-ion system to demonstrate a hardware-efficient QEC implementation that surpasses the break-even point. The logical qubit is encoded in multiple internal Zeeman sublevels of a single trapped ion, and an additional ancilla ion is introduced for dissipation. The dominant noise source in our demonstration, the first-order dephasing error, can be effectively corrected by a multi-level code, and we experimentally realized this code in the $3^2D_{5/2}$ manifold of a $^{40} \rm{Ca}^{+}$ ion. The autonomous QEC protocol is implemented via a tailored Raman sideband transition, which maps the error-induced logical state distortion into phonon excitation. Subsequent sympathetic cooling with an ancilla ion resets the ion motion to the ground state, removing unwanted entropy. 
By applying autonomous QEC repetitively, we observe in the majority cases the logical qubit lifetime is longer than that of the best physical qubit, despite fluctuating ambient dephasing noise. We confirm the QEC performance by adding additional low-frequency noise, where we observe 
substantial longer logical qubit lifetime of $\simeq$11.6 ms comparing with both the physical qubit lifetime ($\simeq$0.9 ms) and the uncorrected logical qubit ($\simeq$0.8 ms), deterministicly surpassing the break-even point of QEC in trapped ion system. Our work establishes a novel pathway toward fault-tolerant quantum computation in trapped-ion systems.

\begin{figure}[tb]
	\begin{center}
		\includegraphics[width=\columnwidth]{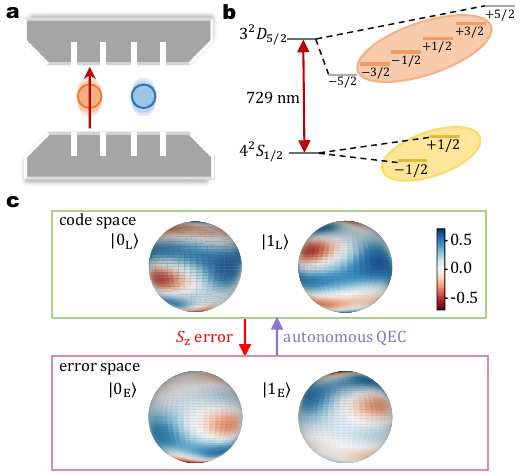}
        \caption{\textbf{Experimental scheme for QEC code with multiple levels.} (a) Schematic of the experimental setup showing two ions ($^{40}\mathrm{Ca}^{+}$) in a linear blade trap: one ion (orange) for encoding the logical qubit in the multiple internal states, and an ancilla ion (blue) for implementing autonomous QEC via sympathetic cooling. (b) Energy level diagram of the ion, highlighting the four-level subspace $\{|\pm3/2\rangle_D,\,|\pm1/2\rangle_D\}$of $3^2D_{5/2}$ manifold for encoding logical code space and error space, with $4^2S_{1/2}$ sublevels serving as the auxiliary levels. (c) Wigner function representations of the encoded logical state $\{|0_{\mathrm{L}}\rangle, |1_{\mathrm{L}}\rangle\}$ (upper panel) and the corresponding error states $\{|0_{\mathrm{E}}\rangle, |1_{\mathrm{E}}\rangle\}$ (lower panel).}
		\label{Fig1}
	\end{center}
\end{figure}

\smallskip{}

\noindent \textbf{\large{}Results}{\large\par}

\noindent The essence of QEC is to encode quantum information in an expanded Hilbert space, utilizing redundancy to detect the potential errors and protect quantum information. Figure~\ref{Fig1}a illustrates our experimental implementation, where a pair of \({}^{40}\mathrm{Ca}^{+}\) ions confined in a linear Paul trap are employed. One ion serves as the carrier to encode quantum information, with its relevant energy level structure depicted in Fig.~\ref{Fig1}b, while the other ion functions as an ancilla for sympathetic cooling that assists the autonoumous QEC process. For Zeeman sublevels in trapped-ion systems, spin flip errors can be substantially suppressed while the phase flip errors ($S_{\mathrm{z}}$, the angular momentum operator along the z-direction) due to magnetic field fluctuations are severe (see Methods). To protect against this dominate dephasing noise channel, we designed a four-level QEC code constructed from the \(3^2D_{5/2}\) manifold of the encoded ion, i.e., the logical state basis is formed by superpositions of four selected magnetic sublevels \(|m_S\rangle_D\) (\(m_S = \pm\frac{1}{2}, \pm\frac{3}{2}\)) as
\begin{eqnarray}
\label{Eq.log_qubit}
|0_{\mathrm{L}}\rangle &\equiv& \sqrt{\frac{1}{4}}|-\frac{3}{2}\rangle_D+\sqrt{\frac{3}{4}}|+\frac{1}{2}\rangle_D, \\
|1_{\mathrm{L}}\rangle &\equiv& \sqrt{\frac{3}{4}}|-\frac{1}{2}\rangle_D+\sqrt{\frac{1}{4}}|+\frac{3}{2}\rangle_D,
\end{eqnarray}
satisfying the Knill-Laflamme conditions~\cite{Nielsen2023} for QEC that the correctable error processes map the code space to mutually orthogonal subspaces (error spaces). When $S_{\mathrm{z}}$ error acts on our logical states, as represented by the Wigner functions of code words shown in Fig.~\ref{Fig1}c, the logical states are projected to the error space spanned by
\begin{eqnarray}
\label{Eq.err_qubit}
|0_{\mathrm{E}}\rangle &\equiv& S_{\mathrm{z}}|0_{\mathrm{L}}\rangle =\sqrt{\frac{3}{4}} |-\frac{3}{2}\rangle_D-\sqrt{\frac{1}{4}}|+\frac{1}{2}\rangle_D, \\
|1_{\mathrm{E}}\rangle &\equiv& S_{\mathrm{z}}|1_{\mathrm{L}}\rangle  =\sqrt{\frac{1}{4}}|-\frac{1}{2}\rangle_D-\sqrt{\frac{3}{4}}|+\frac{3}{2}\rangle_D.
\end{eqnarray}

\begin{figure}[tb]
	\begin{center}
		\includegraphics[width=1\columnwidth]{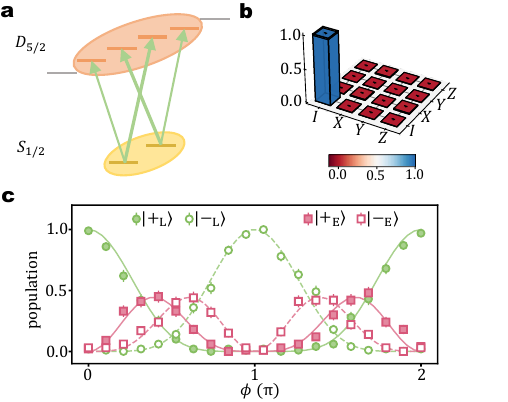}
        \caption{\textbf{Logical qubit encoding and error channel characterization.}
        \textbf{a}, Experimental scheme for encoding and decoding process between the auxiliary ground levels (physical qubit) and the logical qubit. The ``V"-type transition employs state-selective excitation with tailored Rabi frequencies (indicated by line thickness). \textbf{b}, Quantum process ($\chi$) matrices of the combined encoding and decoding process. The height and color of the histogram represent the real parts of the $\chi$ matrix elements. \textbf{c}, Experimental validation of the logical state space evolution with respect to controlled phase rotations. The population evolution of logical space ($|\pm_{\mathrm{L}}\rangle$) and error space ($|\pm_{\mathrm{E}}\rangle$)  basis states are plotted as a function of the rotation angle $\phi$, starting from the initial state $|+_{\mathrm{L}}\rangle$ state. Solid lines represent theoretical predictions, while dots display the experimental results. Each experimental data point was obtained with 100 repetitions, and error bars indicate one standard deviation.
        }
		\label{Fig2}
	\end{center}
\end{figure}

We experimentally implement encoding and decoding of logical states by coherently mapping quantum states between the ground manifold $4^2S_{1/2}$ (auxiliary qubit) and the logical code space $3^2D_{5/2}$ $\{|0_{\mathrm{L}}\rangle,|1_{\mathrm{L}}\rangle\}$. As illustrated in Fig.~\ref{Fig2}a, these processes employ two pairs of phase-coherent 729\,nm resonant laser beams with precisely controlled amplitudes and phases. The performance of the encoding and decoding operations is characterized through quantum process tomography. We prepare the ion in a complete set of initial states in auxiliary qubit (the computational basis states and their superpositions), apply the encoding followed by immediate decoding, and measure the final state in the ground manifold (see Methods for details). The resulting process matrix $\chi_\mathrm{M}$, as shown in Fig.~\ref{Fig2}b, exhibits a process fidelity of 98\%, defined as $F_\chi =\mathrm{Tr}(\chi_\mathrm{M} \chi_\mathrm{I} )$ with $\chi_\mathrm{I}$ being the process matrix for the ideal operation, i.e., identity operation.

To visualize how phase errors affect our encoded logical states, we examine the system's response to controlled phase rotations. We prepare the encoded ion in the superposition state $|+_{\mathrm{L}}\rangle$, and measure the population of the states $|\pm_{\{\mathrm{E},\mathrm{L}\}}\rangle=\frac{1}{\sqrt{2}}(|0_{\{\mathrm{E},\mathrm{L}\}}\rangle\pm|1_{\{\mathrm{E},\mathrm{L}\}}\rangle)$ after applying a controlled rotation $R_{\mathrm{z}}(\phi)=e^{i\phi S_{\mathrm{z}}}$ with varying angles $\phi$. This rotation is implemented by introducing calibrated frequency detunings across the beams, simulating a static magnetic field fluctuations that deviated from a calibrated value.  Figure~\ref{Fig2}b displays the experimental results. When $\phi$ is small, we observe that population primarily transfers to the error state $|+_{\mathrm{E}}\rangle$, consistent with first-order error ($S_{\mathrm{z}}$) behavior predicted by our theoretical model. As $\phi$ increases, the quantum state evolution becomes more complex, with population gradually appearing in the $|-_{\mathrm{E}}\rangle$ and $|-_{\mathrm{L}}\rangle$, indicating increased probability of second- and third-order errors ($S_{\mathrm{z}}^2$, $S_{\mathrm{z}}^3$) that can not be addressed by our QEC code.

\begin{figure*}[tb]
	\begin{center}
		\includegraphics[width=2\columnwidth]{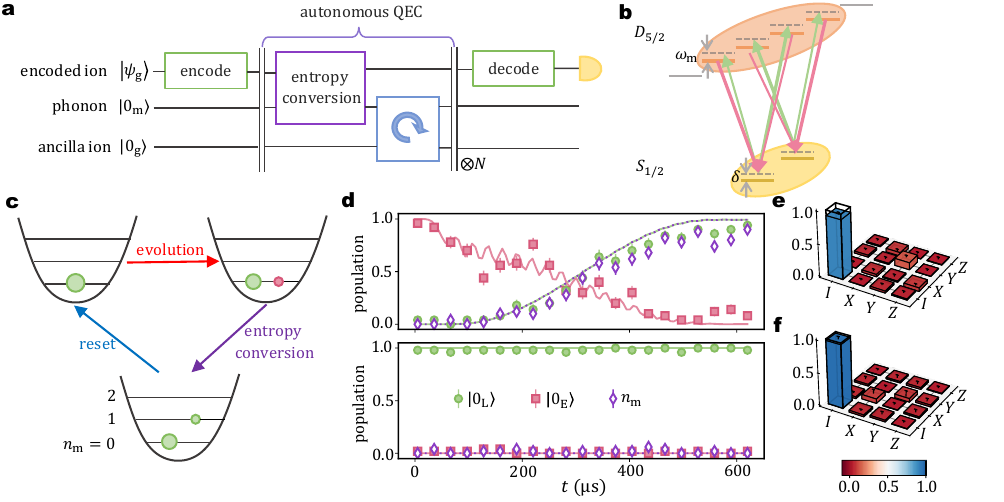}
        \caption{ \textbf{Autonomous QEC procedure.} \textbf{a}, Quantum circuit diagram of the repetitive autonomous QEC. Each repetitions cycle of the autonomous QEC process consists of two steps: the entropy conversion between the logical qubit and phononic mode, and the reset of phonon mode by sympathetic cooling with ancilla ion. (b) Detailed implementation of the entropy conversion mechanism via Raman sideband transitions, where ``V"-type transitions for $|0(1)_\mathrm{g}\rangle\leftrightarrow |0(1)_{\mathrm{L}}\rangle$ (green line) and $|0(1)_\mathrm{g}\rangle \leftrightarrow |0(1)_{\mathrm{E}}\rangle$ (pink line) are simultaneously driven with frequency detunings from auxiliary levels of $-\delta+\omega_\mathrm{m}$ and $-\delta$, respectively, where 
        $\omega_\mathrm{m} = 2\pi \times 1.3\,\mathrm{MHz}$ as the motional frequency, and $\delta=2\pi \times 15\,\mathrm{kHz}$.  \textbf{c}, Conceptual illustration of the autonomous error correction mechanism in the representation of phonon Fock state. An $S_{\mathrm{z}}$ error transforms states in code space (green) into error space (pink), and entropy conversion will coherently maps error space back to code space while  transferring the error-induced entropy to phonon modes, and a sympathetic cooling process will reset the phononic mode back to ground state. \textbf{d}, Population dynamics during the entropy conversion for $|0_{\mathrm{L}}\rangle$ (green circles), $|0_{\mathrm{E}}\rangle$ (pink squares) and phonon excitation (purple diamonds) with the initial state is $|0_{\mathrm{E}}\rangle$ (upper) or $|0_{\mathrm{L}}\rangle$ (bottom). Symbols are experimental results while curves are numerical results. \textbf{e}\&\textbf{f} Quantum process tomography matrices (real components) for a complete autonomous QEC cycle, with the initial state in error space $|0_{\mathrm{E}}\rangle$ (\textbf{e}) or in code space $|0_{\mathrm{L}}\rangle$ (\textbf{f}).
}
		\label{Fig3}
	\end{center}
\end{figure*}

Having established our logical encoding and its performance under phase rotations, we now implement autonomous QEC protocol specifically designed to address $S_{\mathrm{z}}$ error. As illustrated in Fig.~\ref{Fig3}a, our complete QEC scheme consists of three essential components: (1) initial encoding of quantum information into the logical space, (2) repeated cycles of autonomous error correction during the storage period, and (3) final decoding for readout. Each autonomous correction cycle comprises two critical operations: entropy conversion and entropy dissipation. In sharp contrasts to previous demonstrations of QEC, where error syndrome measurement are necessary for tracking errors or implementing feedforward recovery operations, our autonoumous QEC is implemented by applying driving lasers with multiple tones to the ions, as depicted in Fig.~\ref{Fig3}b.

The conceptual foundation of our autonoumous QEC is depicted in Fig.~\ref{Fig3}c. In ideal cases, the internal state of the ion is in the logical space (green circle) and the motional degree of freedom is in the ground state ($\ket{0}_{\mathrm{m}}$), i.e., the lowest phononic Fock energy level in the harmonic potential. During an idling process, the phase error maps the system into an incoherent mixture of the logical space and error space (pink circle), while the motional state is preserved. Then, the entropy conversion operation is implemented through a tailored Raman sideband process by simultaneously applying eight precisely tuned tones of 729\,nm laser radiation, with specific detunings from the ground manifold transitions (see Methods). The process is governed by the effective Hamiltonian:
\begin{equation}
H_{\mathrm{EC}} = \frac{\Omega_{\mathrm{ec}}}{2} ( |0_{\mathrm{L}}\rangle \langle 0_{\mathrm{E}}| + |1_{\mathrm{L}}\rangle \langle 1_{\mathrm{E}}| )\otimes a^{\dagger}+h.c.,
\end{equation}
where $\Omega_{\mathrm{ec}}$ represents the effective Raman coupling strength, $a^{\dagger}$ is the creation operation for the selected motional mode, and $h.c.$ is short for hermitian conjugate. Although the Hamiltonian generates the unitary operation $U=\exp(iHt)$ and allows the inter-conversion between the logical space and code space, only the unidirectional mapping from error space to code space can be implemented given the motion initialized at the ground state. Explicitly, we have
\begin{equation}
    U\ket{0_{\mathrm{E}}/1_{\mathrm{E}}}\otimes \ket{0}_{\mathrm{m}}=\ket{0_{\mathrm{L}}/1_{\mathrm{L}}}\otimes \ket{1}_{\mathrm{m}}
\end{equation}
for $t=\pi/\Omega_{\mathrm{ec}}$, while the reversal process driving out of the $\ket{0_{\mathrm{L}}/1_{\mathrm{L}}}\otimes \ket{0}_{\mathrm{m}}$ states is inhibited as $a\ket{0}_{\mathrm{m}}=0$, leaving the logical space unaffected. Therefore, the entropy of the system due to phase error is transferred to the mixed state of the motional degree of freedom. Figure~\ref{Fig3}d experimentally verifies this entropy conversion process. The upper panel shows the dynamics when starting in the error state $\ket{0_{\mathrm{E}}}\otimes\ket{0}_{\mathrm{m}}$, and a Rabi-like oscillation to $\ket{0_{\mathrm{L}}}\otimes\ket{1}_{\mathrm{m}}$ confirms the unitary nature of the operation, with the optimal entropy transfer occurs at a duration of $620\,\mathrm{\mu s}$. The low panel demonstrates that when starting in the logical state $\ket{0_{\mathrm{L}}}\otimes\ket{0}_{\mathrm{m}}$, the state remains stable. 


To complete the autonomous QEC cycle, as shown in Figs.~\ref{Fig3}a and c, we reset the motional degree of freedom to ground state and remove the entropy through subsequent sympathetic cooling mediated by the ancilla ion. Using tailored 729\,nm laser pulses focused exclusively on the ancilla ion, combined with global 397\,nm laser illumination, the shared motional mode between the two ions can be cooled. It is worth noting that this autonomous QEC operation satisfies the condition of ``error transparency" as introduced in recent experimental works in superconducing quantum platforms~\cite{may2020,maw2020} (see Methods), indicating the potential of our approach in fault-tolerant quantum operations.  

We characterized the performance of the complete autonomous QEC cycle through quantum process tomography, measuring its effect on both logical and error states. Figure~\ref{Fig3}e shows the process matrix ($\chi_\mathrm{M}$) when the initial state is in the error space, while Fig.~\ref{Fig3}f shows the process matrix for an initial state in the code space. The corresponding process fidelity are $F_{\chi}=0.92(2)$ and $F_{\chi}=0.98(1)$, respectively. The slightly reduced fidelity for error space is attributed to imperfections in the entropy conversion and reset processes.

\begin{figure}[tb]
	\begin{center}
		\includegraphics[width=1\columnwidth]{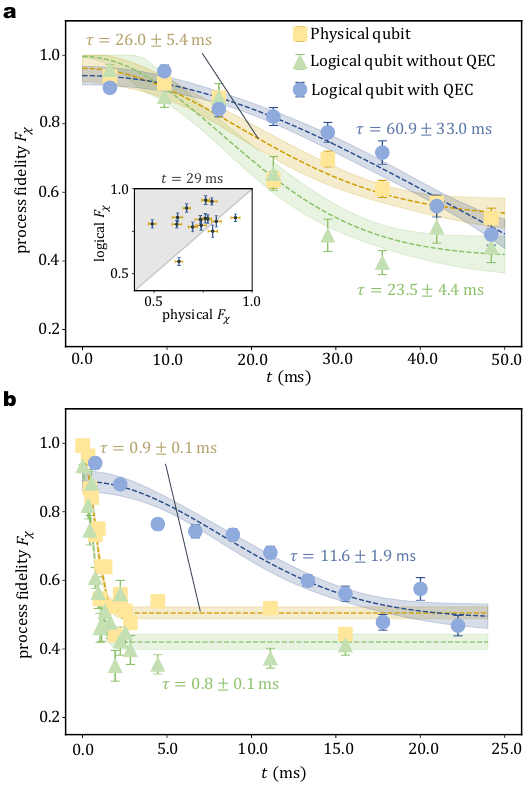}
        \caption{ \textbf{Demonstration of surpassing the break-even point with autonomous QEC.} Process fidelity ($F_{\chi}$) as a function of total evolution time time $t$ under (\textbf{a}) ambient laboratory conditions and (\textbf{b}) injected low-frequency magnetic field noise (cutoff frequency of 10\,Hz). $F_{\chi}$ are measured for optimal physical qubit (yellow squares), encoded qubit without error correction (green triangles), encoded qubit with repetitive autonomous QEC (blue circles). The dashed line represents the fitted curve, and the shaded region corresponds to the fitting error. The inset of \textbf{a} provides a direct statistical comparison between the physical and logical qubit fidelities at $t = 29\,\mathrm{ms}$, i.e., 9 cycles of autonomous QEC, for multiple experimental trials, with the diagonal line indicating the break-even criteria. Each data point consists of at least 100 repetitions, with error bars derived from bootstrap resampling. 
}		
\label{Fig4}
	\end{center}
\end{figure}

By repetitively implement the autonomous QEC procedure, it is anticipated that the coherence time of quantum information can be extended. A critical metric for evluation the performance of QEC is the $\Lambda$-factor~\cite{Acharya2025}
\begin{equation}
    \Lambda=\frac{\tau_{\mathrm{logical}}}{\tau_{\mathrm{physical}}}
\end{equation}
which is defined as the ratio of the logical qubit lifetime ($\tau_{\mathrm{logical}}$) to the physical qubit ($\tau_{\mathrm{physical}}$) lifetime. $\Lambda>1$ means the QEC surpassing the break-even point, indicating the logical qubit outperforms the best available physical qubit consisting the logical qubit~\cite{Ofek2016, Ni2023, Sivak2023, Acharya2025}. To evaluate our logical qubit, the quantum process fidelity decay under three distinct configurations are characterized, and the results are summarized in Fig.~\ref{Fig4}a. The encoded logical qubit with (blue dots) and without (green dots) repetitive autonomous QEC are measured with systematically varied total evolution time $t$. Each autonomous QEC cycle has a total duration $\tau_{\mathrm{AQEC}}=3.2\,\mathrm{ms}$, with a duration of $\tau_{\mathrm{EC}}=0.62\,\mathrm{ms}$ for entropy conversion, and a duration of $\tau_{\mathrm{i}}=2.6\,\mathrm{ms}$ for resetting and idling. As a comparison, the optimal physical qubit encoded in the $\mathrm{span}\{ |-\frac{1}{2}\rangle_D,\,|\frac{1}{2}\rangle_D \}$ is measured (yellow dots). The auxiliary qubit also participates the encoding and decoding processes, while its coherent time is shorter as it is more sensitive to magnetic field fluctuations. We model the temporal decay of process fidelity using Gaussian function $F_{\chi}(t)=Ae^{-(t/\tau)^2}+C$, where $A$ and $C$ are fitting parameters, and $\tau$ corresponds to the lifetime. This Gaussian decay model, rather than an exponential one, better captures the behavior of our system where the dominant noise source is low-frequency magnetic field fluctuations (see Methods). 

Under ambient conditions, the QEC-protected logical qubit demonstrates a fitted lifetime of $\tau_{\mathrm{logical}}=60.9\pm33.0\,\mathrm{ms}$, compared to the physical qubit lifetime of $\tau_{\mathrm{physical}}=26.0\pm5.4\,\mathrm{ms}$, yielding an enhancement factor of $\Lambda=2.3\pm1.4$.
While this indicates improvement, the large uncertainty stems from substantial ambient noise fluctuations in our laboratory environment. Nevertheless, the logical qubit with QEC maintains consistently higher fidelity than both the physical qubit and the uncorrected logical qubit throughout the 20-40 ms time interval, as shown in Fig.~\ref{Fig4}a. The inset of Fig.~\ref{Fig4}a provides a statistical comparison at $t=29\,\mathrm{ms}$ (corresponding to 9 cycles of autonomous QEC), showing that the majority of repetitive QEC trials achieve higher fidelity than the physical qubit trials, further supporting the enhancement of logical qubit lifetime by autonomous QEC.

To more conclusively demonstrate the performance advantage of our QEC protocol, we further conducted experiments with deliberately injected low-frequency magnetic field noise (cutoff frequency of 10\,Hz, see Methods for details). This controlled noise environment provides a more challenging test case to unveil the potential of our logical qubit. The corresponding experimental results are shown in Fig.~\ref{Fig4}b, with a reduced idling time between QEC cycles ($\tau_\mathrm{i}=0.12$ ms) to suppress uncorrectable high-order phase errors. Under the enhanced noise conditions, the differences between the three configurations become dramatically more pronounced, with fitted lifetimes $\tau_{\mathrm{logical}}=11.6\pm1.9\,\mathrm{ms}$ and $\tau_{\mathrm{physical}}=0.9\pm0.1\,\mathrm{ms}$, yielding $\Lambda=12.9\pm2.6$. Both the ambient and enhanced noise conditions validate the advantages of autonomous QEC for protecting quantum information in a single ion against phase errors. Importantly, this improvement is achieved without any measurement~\cite{Ofek2016, Ni2023, Sivak2023, Acharya2025} or post-selection~\cite{Hong2024,Egan2021}, representing a truly deterministic extension of quantum coherence through quantum technology.

\smallskip{}

\noindent \textbf{\large{}Discussion}{\large\par}

\noindent We have demonstrated a novel QEC protocol combining multi-level qudit encoding with autonomous QEC in a trapped-ion platform. Our experimental results conclusively show that this approach significantly extends quantum information lifetime beyond what is achievable with optimal physical qubit encoding, particularly under challenging noise environments. The demonstrated enhancement factor $\Lambda\approx 12.9$ under controlled noise conditions establishes that our protocol decisively surpasses the break-even point of QEC. The autonomous QEC protocol offers several distinct advantages compared to conventional error-syndrome measurement implementations. First, by encoding the logical qubit within a single ion, we avoid the need for complex multi-qubit entangling operations that typically limit the fidelity of the encoding process. Second, our measurement-free approach eliminates the technical challenges associated with mid-circuit detection and real-time feedback, which remain significant hurdles for trapped ion quantum computing. Third, the error conversion mechanism, which transforms logical errors into phonon excitations that can be dissipated through sympathetic cooling, provides a natural and efficient entropy management that leverages the well-established laser-cooling techniques.

While our current implementation focuses on protecting against first-order dephasing errors in a single trapped ion, there are multiple pathways for extending this approach. On one hand, the multi-level encoding demonstrated here is readily extensible for larger Hilbert spaces by incorporating additional Zeeman sublevels, which allows the protection against multiple error types and higher-order errors by increasing the redundancy. On the other hand, the approach can be scaled up to multiple logical qubit by introducing more ions, with independent entropy conversion and reset channels available as each ion introduces additional motional degree of freedom. Future work should explore how to implement entangling operations between logical qubits encoded in different ions while maintaining their error correction properties, and then concatenated codes can be implemented based on logical qubits to achieve higher levels of error suppression~\cite{Putterman2025}. Our approach may find applications beyond trapped-ion systems, extending to other quantum platforms with rich internal structures, such as neutral atoms, solid-state defect centers, or rare-earth ions in crystals. Additionally, the autonomous QEC may also inspire new perspectives on quantum metrology~\cite{Kwon2025} and relevant applications.

\smallskip{}

\noindent \textbf{NOTE}: When finalizing this work, we became aware of a closely related work \cite{Debry2025}, which experimentally demonstrated a similar encoding scheme in a single ion, achieving a logical qubit lifetime that exceeded their ground-state manifold (auxiliary qubit) encoded qubit by a factor of 1.5.

\smallskip{}

\clearpage{}

\clearpage{}
\setcounter{figure}{0}
\renewcommand{\thefigure}{S\arabic{figure}}  
\setcounter{table}{0}
\renewcommand{\thetable}{S\arabic{table}}    
\setcounter{equation}{0}
\renewcommand{\theequation}{S\arabic{equation}} 

\renewcommand{\theHfigure}{S.\arabic{figure}} 
\renewcommand{\theHtable}{S.\arabic{table}}
\renewcommand{\theHequation}{S.\arabic{equation}}

\noindent \textbf{\large{}Supplemental Materials}{\large\par}

\noindent \textbf{Device and setup}

\noindent The experiment is performed using two $^{40}\rm{Ca}^+$ ions in a cryogenic Paul trap \cite{leibfried2003}, with the ambient magnetic field strength being $B=0.9$ mT, provided by a pair of permanent magnets. We utilize the radial rocking mode for error correction, which has a frequency of $\omega_\mathrm{m}=2\pi\times 1.3$ MHz and a heating rate of less than 1 phonon/s. We estimate a Lamb-Dicke parameter of approximately 0.056 for the spin-motion coupling. Doppler cooling and EIT cooling are implemented using 397 nm and 866 nm lasers, followed by sideband cooling, which further reduces the phonon occupation to about 0.02.

Coherent operations are performed via the quadrupole transition between $S_{1/2}$ and $D_{5/2}$ manifolds, driven by a narrow-linewidth 729 nm laser \cite{Ringbauer2022}. Individual addressing is achieved via two crossed acousto-optic deflectors (AODs) controlling the radial 729 nm beam, with a crosstalk Rabi frequency of less than 2\%, which is sufficiently low for our experiment. Thus, the focused 729~nm laser beam can be steered to either of the ions by $\rm{20~\mu s}$, followed by individual operations. To achieve multi-tones 729 nm driving, we apply multiple frequency components, with independently controlled amplitudes and phases, to a single-pass acousto-optic modulator (AOM) prior to the crossed AODs above. State readout is performed using fluorescence detection with 397 nm and 866 nm lasers, and an electron-multiplying charge-coupled device (EMCCD) camera is used for site-resolved detection, with readout fidelity of 99.7\% for each ion. 

\vbox{}

\noindent \textbf{State preparation and measurement} 

\noindent We initialize the ion into $\ket{-1/2}_S$ state of the $S_{1/2}$ manifold using 397~nm $\sigma_-$ laser and 866 nm repump laser and further purified through 729 nm laser resonantly driving transition from $\ket{1/2}_S$ to the  $D_{5/2}$ manifold, combined with 854 nm laser repumping back to the $S_{1/2}$ manifold. The initial unencoded qubit state is prepared using two 729 nm pulses, shown in (see Fig.~S1b), $\pi$ time for each pulse is 20 $\mu s$. The subsequent encoding uses 729 nm lasers with four resonant tones for a pulse duration of 50 $\mu s$. The decoding and readout process is the reverse of such a process, and we use 729 nm shelving that transferring $\ket{1/2}_S\rightarrow\ket{5/2}_D$ to distinguish the $\ket{\pm1/2}_S$ states. The optimal physical qubit states, in the space of $\ket{1/2}_D$ and $\ket{-1/2}_D$ in the $D_{5/2}$ manifold, can also be prepared using two 729~nm laser pulses sequentially driving $\ket{-1/2}_S\leftrightarrow\ket{\pm1/2}_D$ (see Extended Data Fig.~1c).  

\vbox{}

\noindent \textbf{Error correction pulse}

\noindent 
We utilize the auxiliary levels $\ket{\pm1/2}_S$ to implement a Raman sideband transition that transfers the error state $\ket{\psi_{\mathrm{E}}}\otimes\ket{0}_{\mathrm{m}}$ to the logical state $\ket{\psi_{\mathrm{L}}}\otimes\ket{1}_{\mathrm{m}}$. The detuning from $\ket{\pm1/2}_S$ is 15 kHz, and a ramping time of 120 $\mu s$ with sine-squared pulse shape, which is used to suppress off-resonant excitations. The time evolution of this process is shown in Fig.~3 of the main text, with the total time being 620 $\mu s$.

 We choose a direct Raman transition instead of using a two-step method, where a resonant sideband transition first moves the population to the $\ket{\pm1/2}_S$ states before transferring to the logical state $\ket{\psi_{\mathrm{L}}}$. This two-step approach is fundamentally limited by the finite phase coherence time of the $\ket{\pm1/2}_S$ ground states. In contrast, our method avoids direct population transfer to the ground states, which fulfills the requirements for error correction.

\vbox{}
\begin{figure}[]
	\begin{center}
		\includegraphics[width=1\columnwidth]{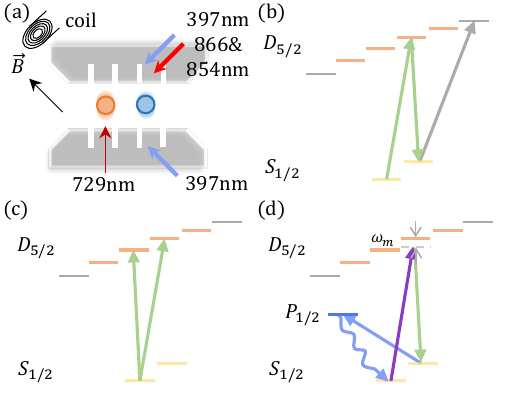}
        \caption{\textbf{Experimental apparatus, state preparation and sympathetic cooling process.} (a) Experimental setup, including the laser beams and magnetic coil for injecting noise. 397, 854, 866 nm are global beams for cooling, repumping and detection. We use an addressed 729 nm laser along the radial direction to distinguish the encoding and ancilla ion.
        (b) We use two 729 nm on-resonance pulses (green arrows) to prepare the state in the ancilla levels, and use electronic shelving (grey arrow) that transfers $|1/2\rangle_S$ to the dark state $|5/2\rangle_D$ for detection. 
        (c) Optimal physical qubit preparation. 
        (d) Sympathetic cooling process using an ancilla ion. 
        }
		\label{figS1}
	\end{center}
\end{figure}
\noindent \textbf{Sympathetic cooling}

\noindent We use an ancilla $^{40}\rm{Ca}^+$ ion for sympathetic cooling. Starting from the $\ket{-1/2}_S$ state and an excited motional state $\ket{1}_{\mathrm{m}}$, a focused 729~nm laser is applied to drive the spin-motion coupled ``red-sideband" transition of $\ket{-1/2}_S\otimes\ket{1}_{\mathrm{m}}\leftrightarrow \ket{1/2}_D\otimes\ket{0}_{\mathrm{m}}$, followed by another 729~nm laser pulse driving motion preserving ``carrier" spin transition $\ket{1/2}_D\rightarrow\ket{1/2}_S$. As such, a global 397 nm $\sigma_-$ pumping laser and 866~nm laser optically pump the state from $\ket{1/2}_S$ back to $\ket{-1/2}_S$, completing the cycle while removing a quantum of motional excitation. This process effectively removes entropy from the system and is ready for the next round of cooling (see Fig.~S1d). The total time for the cooling operations is 90 $\mu s$, which is much shorter than the coherence time.  This method avoids using the global 854~nm laser for optical pumping during the ground state cooling, which pumps the $D_{5/2}$ manifold to the $S_{1/2}$ manifold and destroys the information stored in the logic ion.

\vbox{}

\noindent \textbf{Process tomography}

\noindent Process tomography is used to benchmark our operation performance. It is realized by preparing four different initial states $\ket{0}, \ket{1}, 1/\sqrt{2}(\ket{0}+\ket{1}), 1/\sqrt{2}(\ket{0}+i\ket{1})$ and performing the corresponding state tomography. This is carried out using the maximum likelihood estimation method. In cases where population leakage occurs outside the computational subspace, we compensate by adding a proportion of the maximally mixed state, ensuring the reconstructed density matrices remain properly normalized.
The process tomography matrices are then derived and the fidelity is defined as $F_\chi =\mathrm{Tr}(\chi_\mathrm{M} \chi_\mathrm{I} )$, with $\chi_\mathrm{M} (\chi_\mathrm{I} )$ being the derived 4 × 4 process matrix for experimental (ideal) operation. 

\vbox{}

\noindent \textbf{Magnetic field fluctuation and injection of noise}

\noindent Our setup is equipped with two layers of magnetic shielding. However, we observe residual magnetic field fluctuations attributed to the nearby elevators, metal ladders, electronic noise in the power supply line and temperature-induced magnetic field drift. Under realistic experimental conditions at midnight, we estimate peak-to-peak magnetic field noise is approximately 0.9 nT on average. However, occasional spikes in the noise could be observed, which may cause large enough errors in the higher order.

To demonstrate the effectiveness of QEC, we inject magnetic noise into the system by applying electronic currents with amplitudes varied from shot-to-shot to the magnetic coils near the trap inside the shielded areas. The currents are randomly sampled from a Gaussian distribution and updated every 100 ms, effectively simulating white noise with a cutoff frequency of 10 Hz. This approach mimics the low-frequency magnetic noise typically present in our system but with a fixed distribution. The standard deviation of the applied magnetic field is about 16 nT, dominant over the above mentioned residual noise.

\vbox{}

\noindent \textbf{QEC code for Lindblad error}

\noindent Here we demonstrate the quantum error correction code for both Lindblad noise and coherent random rotation noise.  Firstly, we calculate the Lindblad noise
\begin{equation}
\frac{d\rho}{dt}=L(\rho)=\kappa(S_{\mathrm{z}}\rho S_{\mathrm{z}}-\frac{1}{2}S_{\mathrm{z}}^{2}\rho-\frac{1}{2}\rho S_{\mathrm{z}}^{2}),
\end{equation}
which is commonly used to represent a white noise, $\kappa$ is the dephasing rate. The Liouvillian can be represented as a Kraus operator representation $\rho(t)=\sum_{l=0}^{+\infty} E_{\mathrm{L}}(t) \rho(0) E_{\mathrm{L}}(t)^\dagger$, in which $E_{l}=\sqrt{\frac{\kappa^{l}t^{l}}{l!}}e^{-\kappa t/2S_{\mathrm{z}}^{2}}S_{\mathrm{z}}^{l}$ is the $l$ jumps Kraus operator. When $\kappa t\ll 1$, we only calculate the first order error, i.e. $E_0=e^{-\kappa t /2 S_{\mathrm{z}}^2}$, $E_1 = \sqrt{\kappa t}e^{-\kappa t /2 S_{\mathrm{z}}^2} S_{\mathrm{z}}$. To satisfy the Knill-Laflamme condition 
\begin{equation}
\langle\sigma_{\mathrm{L}}|E_{l}^{\dagger}E_{k}|\sigma_{\mathrm{L}}^{'}\rangle=\alpha_{lk}\delta_{\sigma\sigma'},
\end{equation}
the minimum multi-level code can correct errors  $\{ E_0, E_1 \}$ is defined in a four-level subspace $\{|\pm\frac{3}{2}\rangle,\,|\pm\frac{1}{2}\rangle\}$, and the codewords can be written as
\begin{align}
|0_{\mathrm{L}}\rangle&=	\sqrt{\frac{e^{2\kappa t}}{3+e^{2\kappa t}}}|-\frac{3}{2}\rangle
	+\sqrt{\frac{3}{3+e^{2\kappa t}}}|+\frac{1}{2}\rangle,\\
|1_{\mathrm{L}}\rangle&=	\sqrt{\frac{3}{3+e^{2\kappa t}}}|-\frac{1}{2}\rangle
	+\sqrt{\frac{e^{2\kappa t}}{3+e^{2\kappa t}}}|+\frac{3}{2}\rangle. 
\end{align}
We emphasize that this subspace can be found when the total spin angular momentum quantum number $S\geq 3/2$ and $S$ is a half-integer values. On the case $\kappa \tau \ll 1$, the Kraus operators can be simplified as $E_0 \simeq I$, $E_1 \simeq S_{\mathrm{z}}$, we can simplify the code and error space codewords as (1-4) in the main text. Here, a parity operator $P=(S_{\mathrm{x}}^{2}-\frac{5}{4}I)$ is defined to distinguish the logical space and error space, i.e., $P|0_{\mathrm{L}}\rangle=P|1_{\mathrm{L}}\rangle=1$, $P|0_{\mathrm{E}}\rangle=P|1_{\mathrm{E}}\rangle=-1$. 

\vbox{}

\noindent \textbf{Coherent random rotation error}

\noindent Next, we will verify that the aforementioned QEC code can also be used to correct coherent random rotation error, i.e. $U(\theta)=e^{-i\theta S_{\mathrm{z}}}$, $\theta$ is a random phase, which is commonly caused by magnet field fluctuation in experiment. The average effect of random rotation
\begin{equation}
\bar{\rho} = \int f(\theta) U(\theta) \rho U(\theta)^{\dagger} d\theta,
\end{equation}
usually $f(\theta)$ is a probability distribution of $\theta$. After an arbitrary rotation with $\theta\ll 1$, the logical state $|\psi_{\mathrm{L}}\rangle=\alpha|0_{\mathrm{L}}\rangle+\beta|1_{\mathrm{L}}\rangle$ is transformed to
\begin{align}
\nonumber
U(\theta)|\psi_{\mathrm{L}}\rangle \simeq &|\psi_{\mathrm{L}}\rangle+(i\theta)\frac{\sqrt{3}}{2}|\psi_{\mathrm{E}}\rangle\\\nonumber
&+(i\theta)^2(\frac{3}{8}|\psi_{\mathrm{L}}\rangle+\frac{\sqrt{3}}{4}Z_{\mathrm{E}}|\psi_{\mathrm{E}}\rangle) +\mathcal{O}(\theta^3).   
\end{align}
Here, $Z_{\mathrm{E}}$ is the Pauli Z operator in the error space and it will cause a logical error after the error correction operation. Therefore, this QEC code can suppress the first order coherent error.

\vbox{}

\noindent \textbf{Recovery channel}

\noindent  When the errors are within the error-correctable range, the recovery channel $\mathcal{R}(\rho')=R_{0}\rho'R_{0}^{\dagger}+R_{1}\rho'R_{1}^{\dagger}$ can transfer the quantum information from the error space back to the code space , in which $R_{0}=|0_{\mathrm{L}}\rangle\langle0_{\mathrm{L}}|+|1_{\mathrm{L}}\rangle\langle1_{\mathrm{L}}|$, $R_{1}=|0_{\mathrm{L}}\rangle\langle{0}_{\mathrm{E}}|+|1_{\mathrm{L}}\rangle\langle{1}_{\mathrm{E}}|$. 
For a logical qubit happens to the error $\rho' = E_0\rho E_0^{\dagger} +E_1 \rho E_1^{\dagger}$, we implement the recovery channel and the final state can be written as
\begin{equation}
\mathcal{R}(\rho')\propto\rho+\mathcal{O}(\kappa^{2}t^{2},\,\theta^2).
\end{equation}
We emphasize that our code can correct only the first order error, for a logical state $|\psi_{\mathrm{L}}\rangle$, if a second order error happened, i.e.
\begin{equation}
S_{\mathrm{z}}^{2}|\psi_{\mathrm{L}}\rangle\propto\frac{3}{4}(\alpha|0_{\mathrm{L}}\rangle+\beta|1_{\mathrm{L}}\rangle)-\frac{\sqrt{3}}{2}(\alpha|0_{\mathrm{E}}\rangle-\beta|1_{\mathrm{E}}\rangle),
\end{equation}
and the final state after recovery channel
\begin{equation}
\mathcal{R}(S_{\mathrm{z}}^2|\psi_{\mathrm{L}}\rangle)=\frac{3}{7}|\psi_{\mathrm{L}}\rangle\langle\psi_{\mathrm{L}}|+\frac{4}{7}Z_{\mathrm{L}}|\psi_{\mathrm{L}}\rangle\langle\psi_{\mathrm{L}}|Z_{\mathrm{L}},
\end{equation}
which happens a logical $Z_{\mathrm{L}}$ error. 

\vbox{}

\noindent \textbf{Entropy conversion}

\noindent The entropy conversion operation is implemented as Fig.~3b in the main text. We implement an entropy conversion operation to convert errors from the encoded system to the phonon, and damp the phonon by the sympathetic cooling introduced by the ancilla ion. The Hamiltonian of the entropy conversion operation can be written as
\begin{align}
H=&	\sum_{l}\omega_{l}a_{l}^{\dagger}a_{l}-\delta_{0}|0_\mathrm{g}\rangle\langle0_\mathrm{g}|-\delta_{1}|1_\mathrm{g}\rangle\langle1_\mathrm{g}|\\ \nonumber
	&+[\frac{\Omega_{0\mathrm{L}}}{2}e^{i(kz_{i}-\omega_{0L})}|0_{\mathrm{L}}\rangle\langle0_\mathrm{g}|+\frac{\Omega_{0\mathrm{E}}}{2}e^{i(kz_{i}-\omega_{0\mathrm{E}})}|0_{\mathrm{E}}\rangle\langle0_\mathrm{g}|\\ \nonumber	&\;\; +\frac{\Omega_{1\mathrm{L}}}{2}e^{i(kz_{i}-\omega_{1\mathrm{L}})}|1_{\mathrm{L}}\rangle\langle1_\mathrm{g}|+\frac{\Omega_{1\mathrm{E}}}{2}e^{i(kz_{i}-\omega_{1\mathrm{E}})}|1_{\mathrm{E}}\rangle\langle1_\mathrm{g}|\\ \nonumber 
& \;\; +h.c.] 
\end{align}
here $\omega_l$ is the frequency of $l$th phonon mode, $\delta_{0(1)}$ is the detuning from $|0(1)_\mathrm{g}\rangle$ states, $\Omega_{ab}$ is the Rabi frequency and $k$ is wave-vector of the transition $|a_{b}\rangle \leftrightarrow |a_\mathrm{g}\rangle$, separately, $a={0,1}$, $b={\mathrm{L},\mathrm{E}}$ represents distinct logical space and error space. The laser frequency detuning from the $D_{5/2}$ sublevels are set $\omega_{a\mathrm{L}}\approx\omega_\mathrm{m}$ and $\omega_{a\mathrm{E}}\approx0$, $a={0,1}$, $\omega_\mathrm{m}$ is the lowest radial rocking mode used in the experiment. The radial modes of the ions can be quantized as $z_{i}	=\sum_{l}(D^{T})_{il}\sqrt{\frac{\hbar}{2m\omega_{l}}}(a_{l}+a_{l}^{\dagger}),$ 
$(D^{T})_{il}$ describe the coupling amplitude between $i$th ion and $l$th phonon mode, in the range of $[-1,1]$. The Lamb-Dicke parameter for the spin-motion coupling is $\eta_{i,\mathrm{m}}=k(D^{T})_{i,\mathrm{m}}\sqrt{\frac{\hbar}{2m\omega_{\mathrm{m}}}}$, and $\eta\ll1$ is satisfied for the experiment. Using Magnus expansion and neglecting all the counter-rotating and higher-order terms, the effective Hamiltonian can be written as
\begin{align}
H_{eff}&=\delta'_{0\mathrm{L}}|0_{\mathrm{L}}\rangle\langle0_{\mathrm{L}}|+\delta'_{0\mathrm{E}}|0_{\mathrm{E}}\rangle\langle0_{\mathrm{E}}|\\ \nonumber
&+\delta'_{1\mathrm{L}}|1_{\mathrm{L}}\rangle\langle1_{\mathrm{L}}|+\delta'_{1\mathrm{E}}|1_{\mathrm{E}}\rangle\langle1_{\mathrm{E}}|\\ \nonumber
&+[\frac{i\Omega_{0}}{2}|0_{\mathrm{L}}\rangle\langle0_{\mathrm{E}}|a^\dagger +\frac{i\Omega_{1}}{2}|1_{\mathrm{L}}\rangle\langle1_{\mathrm{E}}|a^\dagger\\ \nonumber
&+h.c.],
\end{align}
$\delta'_{a\mathrm{L}}=\omega_\mathrm{m} -\omega_{a\mathrm{L}}+\frac{\Omega_{a\mathrm{L}}^{2}}{4(\omega_{a\mathrm{L}}-\delta_{a})}$ and $\delta'_{a\mathrm{E}}= -\omega_{a\mathrm{E}}+\frac{\Omega_{a\mathrm{E}}^{2}}{4(\omega_{a\mathrm{E}}-\delta_{a})}$ are the detuning of $|a_\mathrm{L}\rangle$ and $|a_\mathrm{E}\rangle$, respectively, $\Omega_{a}=\frac{\eta_{i,\mathrm{m}}\Omega_{a\mathrm{L}}\Omega_{a\mathrm{E}}\omega_\mathrm{m}}{2\delta_{a}(\omega_{a\mathrm{L}}-\delta_{a})}$ is the effective Rabi frequency of the transition $|a_{\mathrm{L}}\rangle$ to $|a_{\mathrm{E}}\rangle$. The phonon sideband is selected when $\delta'_{a\mathrm{L}}-\delta'_{a\mathrm{E}}=0$, and the effective Hamiltonian of entropy conversion becomes

\begin{align}
H_{\mathrm{EC}}=&\frac{i\Omega_{\mathrm{ec}}}{2}[|0_{\mathrm{L}}\rangle\langle0_{\mathrm{E}}|\otimes a^{\dagger}+|1_{\mathrm{L}}\rangle\langle1_{\mathrm{E}}|\otimes a^{\dagger}]\\ \nonumber -&\frac{i\Omega_{\mathrm{ec}}}{2}[|0_{\mathrm{E}}\rangle\langle0_{\mathrm{L}}|\otimes a+|1_{\mathrm{E}}\rangle\langle1_{\mathrm{L}}|\otimes a],
\end{align}
$\Omega_{\mathrm{ec}}=\Omega_{0}=\Omega_{1}$ is the effective Rabi frequency, and the evolution operator of entropy conversion can be written as 
\begin{align}
&\quad U_{\mathrm{EC}} = e^{-iH_{\mathrm{EC}}t}\\ \nonumber
&=	I+[\cos(\frac{\Omega_{\mathrm{ec}} t}{2}\sqrt{a^{\dagger}a})-I](|0_{\mathrm{L}}\rangle\langle0_{\mathrm{L}}|+|1_{\mathrm{L}}\rangle\langle1_{\mathrm{L}}|)\otimes I \\ \nonumber
&+	[\cos(\frac{\Omega_{\mathrm{ec}} t}{2}\sqrt{aa^{\dagger}})-I](|0_{\mathrm{E}}\rangle\langle0_{\mathrm{E}}|+|1_{\mathrm{E}}\rangle\langle1_{\mathrm{E}}|)\otimes I\\ \nonumber
&+	(-i)\sin(\frac{\Omega_{\mathrm{ec}} t}{2}\sqrt{ a^{\dagger}a})\frac{1}{\sqrt{ a^{\dagger}a}}(|0_{\mathrm{L}}\rangle\langle0_{\mathrm{E}}|+|1_{\mathrm{L}}\rangle\langle1_{\mathrm{E}}|) \otimes a^{\dagger}\\ \nonumber
&+	(-i)\sin(\frac{\Omega_{\mathrm{ec}} t}{2}\sqrt{ aa^{\dagger}})\frac{1}{\sqrt{ aa^{\dagger}}}(|0_{\mathrm{E}}\rangle\langle0_{\mathrm{L}}|+|1_{\mathrm{E}}\rangle\langle1_{\mathrm{L}}|)\otimes a. \nonumber
\end{align}
So when the initial phonon state is $|0\rangle_\mathrm{m}$, and sympathetic cooling is applied by the ancilla ion after entropy conversion operation, the encoding system is implemented a quantum operation $\mathcal{R}_{\mathrm{EC}}(\rho)=R_{\mathrm{EC},0}\rho R_{\mathrm{EC},0}^{\dagger}+R_{\mathrm{EC},1}\rho R_{\mathrm{EC},1}^{\dagger}$, in which
\begin{align}
\nonumber
R_{\mathrm{EC},0}&=\langle0|U|0\rangle \\ \nonumber
&=I+[\cos(\frac{\Omega_{\mathrm{ec}}}{2} t)-1][|0_{\mathrm{E}}\rangle\langle0_{\mathrm{E}}|+|1_{\mathrm{E}}\rangle\langle1_{\mathrm{E}}|],\\ \nonumber
R_{\mathrm{EC},1}&=\langle1|U|0\rangle \\ \nonumber
&=-i\sin(\frac{\Omega_{\mathrm{ec}}}{2} t)[|0_{\mathrm{L}}\rangle\langle0_{\mathrm{E}}|+|1_{\mathrm{L}}\rangle\langle1_{\mathrm{E}}|], \nonumber
\end{align}
$\mathcal{R}_{\mathrm{EC}}(\rho) = \mathcal{R} (\rho)$ when $\Omega_{\mathrm{ec}} t = \pi$, is the recovery channel aforementioned. 

\vbox{}

\noindent \textbf{Error transparency}

\noindent Here we emphasize that our entropy conversion operation is error transparent \cite{may2020, maw2020}. An error transparent operation need to satisfy the condition
\begin{equation}
U(T,t)E_{j}U(t,0)|\psi_{\mathrm{L}}(0)\rangle=e^{i\phi(t)}E_{j}U(T,0)|\psi_{\mathrm{L}}(0)\rangle,
\end{equation}
it can be seen as a error happens in the final whenever the error occurs. The Hamiltonian meets the condition of error transparent means 
\begin{equation}
\mathcal{P}_{j}^{\dagger}H(t)\mathcal{P}_{j}=\mathcal{P}_{C}H(t)\mathcal{P}_{C}+c(t)\mathcal{P}_{C},
\end{equation}
here $\mathcal{P}_{\mathrm{C}}=|0_{\mathrm{L}}\rangle\langle0_{\mathrm{L}}|+|1_{\mathrm{L}}\rangle\langle1_{\mathrm{L}}|$ is a projection to the logical space, $\mathcal{P}_{j}=\frac{1}{\sqrt{\alpha_{j}}}E_{j}\mathcal{P}_{\mathrm{C}}$, $\alpha$ is a diagonal real matrix and $\alpha_j$ is the $j$th diagonal element,  $c(t)=-d\phi(t)/dt$. For our entropy conversion operation, $\mathcal{P}_{\mathrm{C}}H_{\mathrm{EC}}\mathcal{P}_{\mathrm{C}}=\mathcal{P}_{\mathrm{z}}^{\dagger}H_{\mathrm{EC}}\mathcal{P}_{\mathrm{z}}=0$, so the entropy conversion operation is an error transparent operation, and the recovery channel satisfies fault-tolerant condition of error correction operation for first order error.

\vbox{}

\noindent \textbf{Analysis of fidelity loss}

\noindent Here we consider a Gaussian probability distribution $f(\theta)=\frac{e^{-\theta^2/2\sigma_{\theta}^2}}{\sigma\sqrt{2\pi}}$, $\sigma_{\theta}$ is the standard variance of the Gaussian distribution, which relies on the time scale of the temporal variation of the noise $\tau_{\mathrm{E}}$. If the time scale of errors $\tau_{\mathrm{E}}$ is much shorter than the total time $t$, the random phase $\theta$ can be seen as the displacement of random walk, and the variance  $\sigma_{\theta}^2 \propto t$; However, since $\tau_{\mathrm{E}}$ is equal to or even larger than t in our experiment, the error can be seen as a quasi-stable error, and the random phase $\theta\propto \delta t$ , $\delta$ is the random frequency fluctuation, so the variance $\sigma_{\theta}^2\propto \sigma_{\delta}^2 t^2$, $\sigma_\delta^2$ is the variance of frequency fluctuation. 

For the optimal physical qubit encoded in the $\{|-\frac{1}{2}\rangle_D,|\frac{1}{2}\rangle_D\}$ subspace,  $S_{\mathrm{z}}=-Z/2$, thus the evolution in this subspace can be denoted as 
\begin{align}
    \epsilon(\rho)&=\int^{\infty}_{-\infty}e^{i\frac{1}{2}\theta Z}\rho e^{-i\frac{1}{2}\theta Z} \frac{e^{-\theta^2/2\sigma_{\theta}^2}}{\sigma\sqrt{2\pi}} d \theta \\\nonumber
    &=\int^{\infty}_{-\infty} \big[\mathrm{cos}^2(\theta/2)\rho+ \mathrm{sin}^2(\theta/2) Z \rho Z \\\nonumber
    &\quad +i\mathrm{sin}(\theta/2)\mathrm{cos}(\theta/2) Z\rho \\ \nonumber
    &\quad  - i\mathrm{sin}(\theta/2)\mathrm{cos}(\theta/2) \rho Z \big]\frac{e^{-\theta^2/2\sigma_{\theta}^2}}{\sigma\sqrt{2\pi}} d \theta.
\end{align}
When evaluating the process fidelity with respect to an identity process, it is equal to the process component $\chi_{II}$, thus $F_{\chi}=\chi_{II}=\int^{\infty}_{-\infty} \mathrm{cos}^2(\theta/2)\frac{e^{-\theta^2/2\sigma_{\theta}^2}}{\sigma\sqrt{2\pi}} d \theta=\frac{1}{2}e^{-\sigma_{\theta}^2/2}+\frac{1}{2}$. Then, for the logical qubit encoded in the $\{|0_{\mathrm{L}}\rangle,\,|1_{\mathrm{L}}\rangle\}$ without the autonomous QEC,  the average fidelity
\begin{align}
\mathcal{F} &= \int_{\psi_{\mathrm{L}} \in V} \int_{-\infty}^{+\infty} f(\theta)|\langle \psi_{\mathrm{L}}| U(\theta) |\psi_{\mathrm{L}}\rangle|^2 dV d\theta  \\ \nonumber
&= \frac{5}{12}+\frac{1}{4}e^{-2\sigma_{\theta}^2}+\frac{5}{16}e^{-\sigma_{\theta}^2/2}+\frac{1}{48}e^{-9\sigma_{\theta}^2/2},
\end{align}
is the superposition of several Gaussian functions, and $\mathcal{F}=\frac{dF_{\chi}+1}{d+1}=\frac{2F_{\chi}+1}{3}$.

For autonomous QEC case,  besides $\tau_{\mathrm{E}}$ and $t$, the time scale of autonomous QEC $\tau_{\mathrm{AQEC}}$ is also needed to be considered,  there are three different cases.  First, when $\tau_{\mathrm{E}}\ll\tau_{\mathrm{AQEC}}$, the variance $\sigma_{\theta}^2 \propto \tau_{\mathrm{AQEC}}$;  Next, for $\tau_{\mathrm{E}} \simeq \tau_{\mathrm{AQEC}}$, here $\sigma_{\theta}^2 \propto \tau_{\mathrm{AQEC}}^2$; For the first two situations, we treat the random phases in each error correction step as mutually independent, such that upon phase averaging, the Kraus operators representation $\epsilon(\rho)=K_0\rho K_0^{\dagger}+K_1 \rho K_1^{\dagger}$ corresponding to a single autonomous QEC step is obtained, here $K_0 = \sqrt{1-p} I_{\mathrm{L}}$, $K_1=\sqrt{p}Z_{\mathrm{L}}$, $p=\frac{1}{2}-\frac{3}{8}e^{-\sigma^2/2}+\frac{1}{16}e^{-9\sigma^2/2}-\frac{3}{16}e^{-\sigma^2/2}$, the average fidelity under n steps autonomous QEC $\mathcal{F}=\frac{2}{3}+\frac{1}{3}(1-2p)^n$, which is exponentially decay with $n$; And when $\tau_{\mathrm{E}} \gg \tau_{\mathrm{AQEC}}$, $\sigma_{\theta}^2 \propto \tau_{\mathrm{AQEC}}^2$, but the decay of fidelity is in a Gaussian envelope with $n$, i.e. 
\begin{align}
\mathcal{F} &\gtrsim \frac{2}{3}+\frac{1}{3} e^{-n^2 \sigma^2/2} [\frac{9}{8}-\frac{1}{8}e^{-4\sigma^2n}]^n \\ \nonumber
&=\frac{2}{3}+\frac{1}{3} [\frac{9}{8}e^{-n \sigma^2/2}-\frac{1}{8}e^{-9n\sigma^2/2}]^n,
\end{align}
here $\frac{9}{8}-\frac{1}{8}e^{-4\sigma^2n}>1$ when $n>1$, thus the decay rate of autonomous QEC case is less than the physical one. For small values of $n\sigma^2$, the decay of $\mathcal{F}$ scales as $n^3$, deviating from the Gaussian dependence $\propto n^2$.

Experimentally, $\tau_{\mathrm{E}} \gg \tau_{\mathrm{AQEC}},\,t$ holds regardless of whether additional noise is applied,  so a Gaussian fitting $\mathcal{F}(t)=Ae^{-(t/\tau)^2}+C$ is implemented for our experimental result, which can be seen in Fig.~\ref{Fig4} of the main text.

\vbox{}

\noindent \textbf{Error budget for QEC}

\noindent Here we present the main error source for one cycle of QEC, both from logical states and error states (see Table S1).

\vbox{}

\noindent \textbf{Phonon mode frequency drift}

\noindent The fidelity of the Raman sideband transition for the error correction operator depends on the precise phonon frequency. In our experiment, the radial mode experiences peak-to-peak fluctuations of approximately 200 Hz due to external temperature drift and stray electric field noise. This frequency instability results in a 2.6\% error when starting from the error states.

\vbox{}

\noindent \textbf{Laser intensity fluctuation}

\noindent We implement the sample and hold technique to reduce the long-term fluctuation of laser intensity. This ensures the average intensity fluctuation remains within 1\% for each experimental sequence. Consequently, there's approximately a 1.7\% degradation in error correction fidelity when starting from error states, while having negligible impact on logical states. 

\vbox{}

\noindent \textbf{AC Stark shift}

\noindent Since we use multi-level to encode quantum information, when the laser interacts with these levels, off-resonant coupling induces level shifts and additional phase accumulation, known as the AC Stark shift \cite{Ringbauer2022}. To mitigate this effect, we optimize the laser polarization to maximize transitions for $\Delta m=\pm 1$ and use a moderate laser intensity. Additionally, we measure the phase shift after one error correction cycle, which is about 0.07 $\pi$ (corresponding to an AC Stark shift of approximately 200 Hz). We compensate for this by adjusting the phase of the readout pulses accordingly. The residual error is 0.75\% starting from logical states and 1.1\% starting from error states. Considering the miscalibration of the phase, which is negligible for logical states and about 0.03 $\pi$ for error states, which leads to an additional 0.34\% error.

\vbox{}

\noindent \textbf{Nonzero phonon mode occupation}

\noindent High-quality ground-state cooling and low phonon heating rate are essential for multiple consecutive error correction cycles. After sideband cooling and each sympathetic cooling process, the phonon number is about 0.02, leading to 1.4\% error from logical states and 0.50\% error from error states.

\vbox{}

\noindent \textbf{Shaped pulse imperfections}

\noindent The error-correction pulse incorporates 120 $\mu s$ sine-squared tapers at the beginning and end of the pulses. These shaped segments suppress population leakage to auxiliary levels during the Raman process, remaining 0.52\% error starting from error states.

\vbox{}

\noindent \textbf{Magnetic field fluctuation}

\noindent Our code is useful only for the first-order phase error. Consequently, if the phase error exceeds the code’s distance, it becomes uncorrectable. Under natural conditions without additional magnetic field noise, we only retain the results where the difference between pre- and post-calibrated magnetic field values is less than 0.9 nT. This results in negligible impact (\textless0.01\%) on error correction starting from both states. When injecting artificial magnetic field noise, this leads to a 0.83\% error starting from logical states and a 26\% error starting from error states. 

\vbox{}

\smallskip{}

\begin{table}[h]
    \centering
    \begin{tabular}{lll}
        \toprule
        \textbf{Error source} & \textbf{infidelity (error states)} & \textbf{infidelity (logical states)} \\
        \midrule
        Total & 6.7\% & 2.2\% \\
        \midrule
        Phonon mode \\frequency drift & 2.6\% & \textless0.01\% \\
        \midrule
        Laser intensity \\fluctuation & 1.7\% & \textless0.01\% \\
        \midrule
        AC stark shift & 1.4\% & 0.8\% \\
        \midrule
        Nonzero phonon\\ mode occupation & 0.5\% & 1.4\% \\
        \midrule
        Shaped Pulses\\ Imperfections & 0.5\% & \textless0.01\% \\
        \midrule
        Magnetic field\\ fluctuation & \textless0.01\% & \textless0.01\% \\
        \bottomrule
    \end{tabular}
    \caption{Error budget for one cycle of quantum error correction under ambient laboratory conditions, starting from error and logical states respectively.}
    \label{tab:error_budget}
\end{table}

\newpage
\noindent \textbf{\large{}Acknowledgment}{\large\par}

\noindent We would like to express our gratitude to Prof. Ya Wang, Prof. John Chiaverini and Dr. Kyle DeBry for helpful communications. This work was funded by the National Natural Science Foundation of China (Grant No.~92165206, 92265210, and 12204052), and Innovation Program for Quantum Science and Technology (Grant No.~2021ZD0301603 and 2021ZD0300203). We also acknowledge the support from Anhui Province Science and Technology Innovation Tackle Plan Project. C.-L.Z. and W.C. was also supported by the National Key R\&D Program (Grant No.~2021YFA1402004), the Fundamental Research Funds for the Central Universities, the Supercomputing Center of USTC and the USTC Center for Micro and Nanoscale Research and Fabrication.

\smallskip{}




\clearpage{}

\end{document}